\documentclass{revtex4}
\usepackage{amsmath}
\usepackage{amssymb}
\usepackage{graphicx}

\begin{document}

\title{Observables and unobservables in a non-associative quantum theory}
\author{Vladimir Dzhunushaliev
\footnote{Senior Associate of the Abdus Salam ICTP}}
\email{dzhun@krsu.edu.kg}
\affiliation{Dept. Phys. and Microel.
Engineer., Kyrgyz-Russian Slavic University, Bishkek, Kievskaya Str.
44, 720021, Kyrgyz Republic}

\begin{abstract}
It is shown that the non-associative operators in a non-associative quantum theory are unobservables. The observable quantity may be presented only by the elements of some associative subalgebra. It is shown that the elements of the associative subalgebra are extended objects that can be similar to strings. It is assumed that the non-associative quantum field theory can be applied to the quantization of strongly interacting fields. The method for obtaining field equations in a non-associative case is given.

\end{abstract}


\maketitle

\section{Introduction}

Non-associativity in physics is a very seldom visitor. In Ref.~\cite{jordan} the attempt was made to obtain a possible generalization of quantum mechanics on any numbers including non-associative numbers: octonions. In Ref.~\cite{okubo1995}, the author applies non-associative algebras to physics. This book covers topics ranging from algebras of observables in quantum mechanics, to angular momentum and octonions, division algebras, triple-linear products and Yang - Baxter equations. The non-associative gauge theoretic  reformulation of Einstein's general relativity theory is also discussed. In Ref.~\cite{baez} one can find the review of mathematical definitions and physical applications for the octonions. The modern applications of the non-associativity in physics are: in Refs. \cite{Grossman}, \cite{Jackiw} it is shown that the requirement that finite translations be associative leads to Dirac's monopole quantization condition; in Ref's \cite{Gogberashvili:2005cp} and \cite{Gogberashvili:2005xb} Dirac's operator and Maxwell's equations are derived in the algebra of split-octonions. 
\par
In this paper we would like to show that the application of the non-associativity in quantum theory leads to the interesting fact: the appearance of extended particles which are similar to strings in string theory. 

\section{Unobservables in a non-associative quantum theory}

The observability of a physical quantity $M$ in quantum mechanics means that it is presented as an operator $\widehat M$ with the following properties:
\begin{itemize}
	\item The eigenvalues of the operator $\widehat M$ 
\begin{equation}
	\widehat M \psi = M \psi
\label{1-10}
\end{equation}
gives us the possible spectrum of $M$ values.
	\item The averaged value of the physical quantity is given as 
\begin{equation}
	\left\langle M \right\rangle = \int \psi^* \widehat M \psi d V .
\label{1-20}
\end{equation}
	\item	The values $M$ from \eqref{1-10} and $\left\langle M \right\rangle$ from \eqref{1-20} are real numbers.
\end{itemize}
The same is valid for quantum field theory with some complexifications.
\par 
Now we would like to show that the non-associativity does not allow us to find 
$\left\langle M \right\rangle$ and $M$ as real numbers. The proofs that 
$\left\langle M \right\rangle$ and $M$ are real numbers are well known in quantum mechanics. At the end of the proof that $M$ is a real number we have 
\begin{equation}
	\int \left( \psi^* \widehat M \psi - \psi \widehat M^* \psi^* \right)d V = 
	0 = \left( M - M^* \right) \int \psi^* \psi dV 
\label{1-30}
\end{equation}
and immediately we see that $M = M^*$. The crucial point here is that on the LHS of Eq.~\eqref{1-30} we have the triple products like $\psi^* \widehat M \psi$. In the non-associative quantum theory this product depends on the rearrangements of brackets. Generally speaking 
\begin{equation}
	\left( \psi^* \widehat M \right) \psi \neq 
	\psi^* \left( \widehat M \psi \right) .
\label{1-40}
\end{equation}
The situation for the proof that the averaged value $\left\langle M \right\rangle$ is a real number is similar. 
\par
The main result of this simple consideration is: \emph{non-associative operators present unobservable quantities.}
\par
At the first sight this statement destroys any attempt to give any physical sense to a non-associative quantum theory. Nevertheless the outlet exists: if the non-associative algebra of quantum field operators has an \emph{associative subalgebra} then these associative operators are \emph{observables}. This observation leads to the remarkable result: the observables in a non-associative quantum field theory is a product of nonobservable quantities. 

\section{Observables in a non-associative quantum field theory}

According to the previous remarks, let us consider an observable quantity $\phi$ in a hypothesized non-associative quantum field theory
\begin{equation}
	\widehat \phi \left( x_1, x_2, \ldots , x_n \right) = 
	\begin{cases}
		\text{ either } & \left[ \left[
			\varphi(x_1) \varphi(x_2)  \right] \ldots \varphi(x_n) 
		\right] 
		\\
			\text{ or } & \left[ 
			\varphi(x_1) \left[ \varphi(x_2)  \ldots \varphi(x_n) 
		\right] \right]
	\end{cases}
\label{2-10}
\end{equation}
here $x_i$ are the Minkowskian coordinates; $\varphi(x_i) \in \mathcal A$, $\mathcal A$ is a non-associative algebra of quantum operators $\varphi(x)$; 
$\phi\left( x_1, x_2, \ldots , x_n \right)$ is an element of an associative subalgebra $A \subset \mathcal A$. 
\par 
Now we would like to consider the physical sense of Eq.~\eqref{2-10}. For simplicity we will consider the product of two non-associative operators
\begin{equation}
	\widehat \phi \left( x_1, x_2 \right) = 
		\varphi(x_1) \varphi(x_2) .
\label{2-20}
\end{equation}
It is necessary to mention that the decomposition \eqref{2-20} is very similar to slave-boson 
decomposition in $t-J$~model of High-T$_c$ superconductivity (for review, see Ref.
\cite{lee}) and spin-charge separation in the non-Abelian gauge theories \cite{Niemi:2005qs} - \cite{Faddeev:2006sw}. Let us consider the (anti)commutator 
\begin{eqnarray}
	\left[ 
		\widehat \phi \left( x_1, x_2 \right) , \widehat \phi \left( x_3, x_4 \right)		
	\right]_\pm
  & \stackrel{def}{=} & \widehat \phi \left( x_1, x_2 \right) \widehat \phi \left( x_3, x_4 \right) \pm 
	\widehat \phi \left( x_3, x_4 \right) \widehat \phi \left( x_1, x_2 \right) = 
\nonumber \\
	&& \left[ \varphi(x_1) \varphi(x_2) \right] \left[ \varphi(x_3) \varphi(x_4) \right] 
	\pm 
	\left[ \varphi(x_3) \varphi(x_4) \right] \left[ \varphi(x_1) \varphi(x_2) \right] .
\label{2-30}
\end{eqnarray}
Using $(\pm)$~associators (which are yet unknown) 
\begin{eqnarray}
	\left\{ \varphi(x_1), \varphi(x_2), \varphi(x_3) \right\}_\pm 
	& \stackrel{def}{=} & 
	\left[ \varphi(x_1) \varphi(x_2) \right] \varphi(x_3) \pm 
	\varphi(x_1) \left[ \varphi(x_2) \varphi(x_3) \right] 
\nonumber \\
	&& = \mathfrak{A} \left( x_1, x_2, x_3 \right) , 
\label{2-40}\\
	\left\{ \left( \varphi(x_1) \varphi(x_2) \right), 
		\varphi(x_3), \phi(x_4) \right\}_\pm 
	& \stackrel{def}{=} & 
	\left[ \left[ \varphi(x_1) \varphi(x_2) \right] \varphi(x_3) \right] \varphi(x_4) 
	\pm 
\nonumber \\
	&&\left[ \varphi(x_1) \varphi(x_2) \right] 
	\left[ \varphi(x_3) \varphi(x_4) \right] = 
	\mathfrak{A}_1 \left( x_1, x_2, x_3, x_4 \right) , 
\label{2-43}\\
	\left\{ \varphi(x_1), \varphi(x_2), 
		\left( \varphi(x_3) \phi(x_4) \right) \right\}_\pm 
	& \stackrel{def}{=} & 
	\left[ \varphi(x_1) \varphi(x_2) \right] 	\left[ \varphi(x_3) \varphi(x_4) \right] 
	\pm \varphi(x_1) \left[ \varphi(x_2) \left[ \varphi(x_3) \varphi(x_4) \right] \right]
\nonumber \\
	&& = 
	\mathfrak{A}_2 \left( x_1, x_2, x_3, x_4 \right) , 
\label{2-46}
\end{eqnarray}
we can calculate the RHS of Eq.~\eqref{2-30} 
\begin{equation}
	\left[ 
		\widehat \phi \left( x_1, x_2 \right) , \widehat \phi \left( x_3, x_4 \right)		
	\right]_\pm
	= F \left( x_1, x_2, x_3, x_4 \right).
\label{2-50}
\end{equation}
Here $\mathfrak{A} \left( x_1, x_2, x_3 \right)$ and 
$\mathfrak{A}_{1,2} \left( x_1, x_2, x_3, x_4 \right)$ are some combinations of $\varphi(x_i)$ operators and real functions $\Delta\left( x_1, x_2, x_3 \right)$. $F \left( x_1, x_2, x_3, x_4 \right)$ is an associative operator but probably $F \left( x_1, x_2, x_3, x_4 \right)$ is a real function of coordinates $x_i$. It is very important to emphasize that the object 
$\widehat \phi \left( x_1, x_2 \right)$ is not decomposable, i.e. we can not observe its components $\varphi(x_{1,2})$ because we have shown above that $\varphi(x_{1,2})$ are unobservable quantities. We only can observe the whole object 
$\widehat \phi \left( x_1, x_2 \right) = \varphi(x_{1}) \varphi(x_{2})$. 
\par 
Now we would like to compare this situation with the propagator of string in string theory. In string theory the propagator is a Veneziano amplitude that is the function of four coordinates (or four impulses in the momentum space). Comparing Eq.~\eqref{2-50} with the calculation of a string propagator we can offer the idea that the RHS of \eqref{2-50} is the propagator of an extended object which can be a string if the $F \left( x_1, x_2, x_3, x_4 \right)$ is equal to the Veneziano amplitude. 
\par 
The big difference between string in string theory and extended object in a non-associative quantum field theory is that in the first case the string coordinates are observable quantities but in the second case the inner structure of a non-associative extended object is unobservable.

\section{Field equations in a non-associative quantum field theory}

In the approach to the quantization of strongly interacting fields presented above we assume that any  operator of strongly interacting fields can be presented as the product that is the generalization of slave-boson decomposition in $t-J$~model of High-T$_c$ superconductivity 
\cite{lee} and spin-charge separation in the non-Abelian gauge theories \cite{Niemi:2005qs} - \cite{Faddeev:2006sw}. The non-associative factors $\varphi(x)$ are distributed in the spacetime and should have dynamical equations defining such distribution. Thus the question arises: what kind of field equations describe the dynamics of non-associative operators $\varphi(x)$ ? Our point of view is that the corresponding equations simply are the field equations of the associative field operators $\widehat \phi(x)$. Then all derivatives can be calculated for the non-associative operator $\varphi(x)$ and in the result we deduce the field equations for the non-associative field operators $\varphi(x)$. 
\par 
Let us, for example, consider non-Abelian gauge theory. In this case we have the decomposition 
\eqref{2-10} (which in nothing else but the generalization of spin-charge separation \cite{Niemi:2005qs}~-~\cite{Faddeev:2006sw}). The field equations are the Yang-Mills equations 
\begin{equation}
	\partial_\nu \widehat F^{a \mu \nu} = 0
\label{3-10}
\end{equation}
here $\widehat F^a_{\mu \nu} = \partial_\mu \widehat A^a_\nu - \partial_\nu \widehat A^a_\mu + g f^{abc} \widehat A^b_\mu \widehat A^c_\nu$ is the field strength operator; $\widehat A^a_\mu$ is the operator of gauge field; $a = 1,2, \dots , n$ is the color index for SU(n) gauge field; $f^{abc}$ are the SU(n) structural constants and $\mu, \nu = 0,1,2,3$ are spacetime indices. The operator $\widehat A^a_\mu$ has the following decomposition 
\begin{equation}
	\widehat A^a_\mu \left( x \right) = 
	\begin{cases}
		\text{ either } & \left[ \left[
			\varphi^a_{i_1}(x) \varphi^{i_1}_{i_2}(x)  \right] \ldots 
			\varphi^{i_n}_\mu(x) 
		\right] 
		\\
			\text{ or } & \left[ 
			\varphi^a_{i_1}(x) \left[ \varphi^{i_1}_{i_2}(x)  \ldots 
			\varphi^{i_n}_\mu(x) 
		\right] \right]
	\end{cases}
\label{3-20}
\end{equation}
here $i_1, i_2, \ldots , i_n$ are inner indexes similar to one in spin-charge separation \cite{Niemi:2005qs}~-~\cite{Faddeev:2006sw}. Comparing with the decomposition\eqref{2-10} the decomposition \eqref{3-20} is given at one point $x = x_1 = x_2 = \ldots = x_n$. Inserting the decomposition \eqref{3-20} in the field strength operator $\widehat F^a_{\mu \nu}$ and afterwards in the Yang-Mills equations \eqref{3-10} we shall receive equations for the  nonassociative operator $\phi(x)$.

\section{Outlook}

To summarize we have shown that if a non-associative algebra of quantum field operators has an associative subalgebra, then the operator of extended particles can be represented similarly to the string representation of elementary particles in string theory. 
\par 
It is necessary to note that the decomposition \eqref{2-10} can be thought of as a variation of the idea about hidden variables in the theory of hidden parameters.

\begin{acknowledgments}
V.D. acknowledges D. Singleton for the invitation to do research at Fresno State University and the support of a CSU Fresno Provost Award Grant. 
\end{acknowledgments}

\end{document}